\documentclass[num-refs]{wiley-article}
\usepackage[utf8]{inputenc}
\usepackage{lipsum}
\usepackage{booktabs}
\PassOptionsToPackage{hyphens}{url}
\usepackage{hyperref}
\usepackage{multirow}
\usepackage{xcolor}
\hypersetup{
    colorlinks,
    linkcolor={red!50!black},
    citecolor={blue!50!black},
    urlcolor={black}
}
\newcommand*{\restartrowcolors}{%
  \ifhmode\unskip\fi
  \vadjust{%
    \global\rownum=0 %
  }%
}
\usepackage{tikz}
\usetikzlibrary{shapes,arrows,positioning}
\usepackage{subcaption}
\usepackage{textcomp}

\usepackage{listings}
\usepackage{lstlangarm}

\usepackage[ruled]{algorithm}
\usepackage[noend]{algpseudocode}
\lstdefinestyle{customcpp}{%
  belowcaptionskip=1\baselineskip,
  breaklines=true,
  xleftmargin=\parindent,
  language=C++,
  showstringspaces=false,
  basicstyle=\linespread{0.4}\footnotesize\ttfamily,
  keywordstyle=\bfseries\color{green!40!black},
  numberstyle=\tiny,
  commentstyle=\itshape\color{purple!40!black},
  identifierstyle=\bfseries\color{black},
  stringstyle=\color{red},
  emph={int,char,double,float,unsigned},
  emphstyle=\color{blue},
  morekeywords={uint64_t,uint32_t,__m256i,__m128i,UINT64_C},
}
\lstdefinestyle{customasmarm}{%
  belowcaptionskip=1\baselineskip,
  breaklines=true,
  xleftmargin=\parindent,
  language=[ARM]{Assembler},
  showstringspaces=false,
  basicstyle=\linespread{0.4}\small\ttfamily,
  keywordstyle=\bfseries\color{green!40!black},
  numberstyle=\tiny,
  commentstyle=\itshape\color{purple!40!black},
  identifierstyle=\bfseries\color{black},
  stringstyle=\color{red},
  emph={int,char,double,float,unsigned},
  emphstyle=\color{blue},
  morekeywords={uint64_t,uint32_t,__m256i,__m128i,UINT64_C},
}

\definecolor{light-gray}{gray}{0.95}

\usepackage{todonotes}
\usepackage{microtype}
\usepackage{CJKutf8}
\usepackage[per-mode=symbol]{siunitx}
\usepackage{pgfplots}
\usepackage{tikz}
\usepackage{tikzscale}
\newcommand{\asciicharacter}[1]{`\texttt{#1}'}

\usetikzlibrary{automata,positioning,arrows,quotes}

\definecolor{bblue}{HTML}{4F81BD}
\definecolor{rred}{HTML}{C0504D}
\definecolor{ggreen}{HTML}{9BBB59}
\definecolor{ppurple}{HTML}{9F4C7C}
\definecolor{ggreen}{HTML}{00FF00}

\title{Scanning HTML at Tens of Gigabytes per Second 
on ARM Processors}
\author[1]{Daniel Lemire}

\affil[1]{Data Science Research Center, Universit\'e du Qu\'ebec (TELUQ), Montreal, Quebec, H2S 3L5, Canada}

\corraddress{Daniel Lemire, Data Science Research Center, Universit\'e du Qu\'ebec (TELUQ), Montreal, Quebec, H2S 3L5, Canada}
\corremail{daniel.lemire@teluq.ca}

\fundinginfo{Natural Sciences and Engineering Research Council of Canada, Grant Number: RGPIN-2024-03787}

\runningauthor{Daniel Lemire}
\begin{document}


\maketitle
\begin{abstract}
Modern processors have instructions to process 16~bytes or more at once. These instructions are called SIMD, for single instruction, multiple data.
Recent advances have leveraged SIMD instructions to  accelerate parsing of common Internet formats such as JSON and base64. 
The two major Web browser engines (WebKit and Blink)  have adopted SIMD algorithms for parsing HTML on 64-bit ARM processors. 
During HTML parsing, they quickly identify specific characters with a strategy called \emph{vectorized classification}. We review their techniques and compare them with a faster alternative. 
We measure a 20-fold performance improvement in HTML scanning compared to traditional methods on recent ARM processors. Our findings highlight the potential of SIMD-based algorithms for optimizing Web browser performance.
\keywords{Text Parsing, Vectorization, Web Performance}
\end{abstract}

\section{Introduction}


SIMD (Single Instruction, Multiple Data) instructions allow operations to be performed on multiple words simultaneously. For example, we can compare sixteen~bytes with sixteen~other bytes with one instruction.
Commodity processors (e.g., Intel, AMD, ARM, POWER) have supported
single-instruction-multiple-data (SIMD) instructions for decades. 
Recent ARM processors have four instruction units capable 
of executing SIMD instructions~\cite{applesilicon} and thirty-two SIMD registers.

Nejati and Balasubramanian found that computational tasks such as HTTP parsing may account for for over 60\% of the page load latency on mobile browsers~\cite{10.1145/2872427.2883014}.
Yet, until recently,  mainstream Web browsers 
did not
use explicitly SIMD instructions for critical tasks such as parsing HTML pages.
Most browsers today use either the WebKit or Blink Web engines. WebKit is developed primarily by Apple for their Safari browser. WebKit is
derived from KHTML, a Web engine developed for Linux and BSD by the KDE project. In turn,
Google based their Blink engine on WebKit. It is now part of Google Chrome, Brave and Microsoft Edge. See Table~\ref{tab:history}. 

Blink and WebKit share many similarities both because of their common origin, but also because they tend to emulate each other. For example, the WebKit engine adopted a fast path for HTML parsing called \texttt{HTMLFastPathParser} first developed for Blink. This fast HTML parser supports only a fixed set of common tags: it is used when JavaScript assigns HTML content to an element.

In May~2024, WebKit adopted SIMDe~\cite{simde}, a framework to help support SIMD instructions 
across various systems.
SIMDe was quickly used 
to accelerate HTML parsing (i.e., the \texttt{HTMLFastPathParser} function).\footnote{\url{https://github.com/WebKit/WebKit/pull/28251}}
Maybe coincidentally, again in May~2024, another major Web engine (Blink) adopted Highway~\cite{highway}, a software framework similar to SIMDe---along with a strikingly similar optimization to the HTML parsing.\footnote{\url{https://chromium-review.googlesource.com/c/chromium/src/+/5538407}} The net result is that---within a few weeks---nearly the same SIMD-based optimization was adopted by the two main Web engines.
The Google engineers reported a 2\% performance boost on the jQuery component of the Speedometer~3 test~\cite{speedometer} following this optimization.

We write Web pages using the Hypertext Markup Language (HTML)~\cite{html}. Thus HTML is probably one of the most
important document formats in software history.
HTML is a text format most commonly 
encoded as UTF-8---a superset of ASCII\@. A fundamental component of HTML
is the element. Typically, an element begins with a starting
tag (e.g., \texttt{<name>}) and ends with a closing tag (e.g., \texttt{</name>}). The ampersand \asciicharacter{\&}  represents \emph{HTML entities}: special sequences that are typically used
to represent special characters. For example,  \asciicharacter{\&lt;} represents
the less-than character, 
\asciicharacter{\&gt;} represents
the greater-than character,   
\asciicharacter{\&quot;} represents the quote. 


\begin{table}[]
    \centering
    \begin{tabular}{cccl}\toprule
     Web engine & vendor & parent & time \\\midrule
      KHTML   & KDE & khtmlw library & 1998--\ldots \\
      WebKit   & Apple & KHTML & 2005--\ldots \\
      Blink   & Google & WebKit & 2013--\ldots \\\bottomrule
    \end{tabular}
    \caption{History of Web Engines}
    \label{tab:history}
\end{table}

Both major Web engines (WebKit, Blink) rely on the same fundamental
step when parsing HTML as part of the HTML tokenization. They scan the text input for the less-than character
\asciicharacter{<} which marks the beginning of a tag. Furthermore,
three other characters must be detected as they trigger special code paths:
\begin{itemize}
    \item Whenever the carriage return character is found, browsers normalize the newlines~\cite{infranormalize}. Specifically, all sequences of the carriage return character followed by the line feed character (\asciicharacter{\textbackslash{}r\textbackslash{}n}) are replaced
by line feed character (\asciicharacter{\textbackslash{}n}), and all remaining carriage return characters are replaced by a line feed character.
\item The ampersand character must be detected as it may indicate the presence of an entity which must
be replaced.
\item We must also detect null characters (\asciicharacter{\textbackslash{}0}) as they trigger an error and a  replacement.
\end{itemize}
In practice, we expect the software to jump from one less-than character \asciicharacter{<} to another, as they are most common.
Fig.~\ref{fig:simple} illustrates how Web engines solve this problem with conventional C code.

\begin{figure}
    \centering
    \begin{tabular}{c}
\begin{lstlisting}[style=customcpp]
// pos is a pointer inside the document
while (pos != end && *pos != '<') {
  if (*pos_ == '&' || *pos_ == '\r' || *pos == '\0') {
    return SPECIAL; // special handling
  }
  ++pos;
}
\end{lstlisting}
    \end{tabular}
    \caption{C code to move a pointer to the next less-than character}
    \label{fig:simple}
\end{figure}

More generally, programmers often write code to scan a string of characters and identify some classes of characters. For example, we might want to find all spaces, and numbers and so forth. The conventional approach is to load the characters one by one and to identify them using comparisons, as in Fig.~\ref{fig:simple}.
It is a sufficiently common task that standard libraries provide support for
such tasks. For example, the C++ standard library offers the
\texttt{std::string::find\_first\_of} function. In C\#, the standard library 
has a powerful \texttt{SearchValues} class. 
Thus, even though the problem solved by the engineers working on major Web engines is specific, it is also an instance of a generic problem of wide interest.

\section{SIMD Algorithm}

To scan text faster, we might want to load blocks of consecutive characters and classify them at once.
We refer to the general approach where one classifies characters using SIMD instructions as \emph{vectorized classification}~\cite{langdale2019parsing}. It comes in
different forms. For example, a single SIMD comparison may be needed: e.g., if we need to compare a single character with every character in the block.
In more complex cases, a sequence of instructions may be required: e.g., when validating Unicode strings~\cite{keiser2021validating,schroder2024validating} or parsing URLs~\cite{nizipli2024parsing}. We might have conditions, e.g., a character type must be followed by another character.

At a high level, the vectorized classification problem recently solved by the Web engines WebKit and Blink is described by Algorithm~\ref{alg:vector_classification}. We repeatedly load blocks of $k$~consecutive characters in a vector, and check whether one of these characters is in the desired set---where $k$ might be 16, 32, 64, \ldots. If not, we advance by $k$~characters and load a new block of $k$~characters. See \S~\ref{sec:webkit}
and \S~\ref{sec:blink} for the WebKit and Blink implementations.

In Algorithm~\ref{alg:vector_classification},
we proceed character-by-character
when we have fewer than $k$~characters left in the document.
Alternatively, we could also load the remaining characters in a partial vector, though it does not make a substantial difference. 


\begin{algorithm}
\caption{\centering\strut---
Vectorized classification to search for the next character in the set \asciicharacter{<}, \asciicharacter{\textbackslash{}r}, \asciicharacter{\&}, \asciicharacter{\textbackslash{}0}.
\label{alg:vector_classification}}
\begin{algorithmic}[1]
\Require String $s$ of length $n$, indexed from 0 to $n-1$
\Require $V$ a vector containing $k$ elements (characters)
\Require $Q$ a vector containing $k$ elements (Boolean)
\State Set $t\leftarrow \{<,\&, \backslash{}r, \backslash{}0\}$
\State $i\leftarrow 0$
\While{$i+k < n$}
\State Load characters at indexes $i, i+1, \ldots, i+k-1$ from string $s$ in vector $V$
\State let $Q_j\leftarrow (V_i \in t)$ for $j= 0, \ldots, k$  \Comment{vectorized classification}
\If{ $Q$ contains at least one true element}
\State \Return $i +$ the index of the first true value in $Q$
\EndIf
\State $i\leftarrow i + k$
\EndWhile
\State scan the remaining characters for $s$ from index $i$ to $n-1$ for characters in the set, if found \Return the index
\State \Return that the string does not contain characters from the set
\smallskip
\end{algorithmic}
\end{algorithm}

Algorithm~\ref{alg:vector_classification} might be suboptimal. Consider an example. Suppose that we have
a string that begins with `\texttt{<!doctype html><html itemscope="" itemtype="http://schema.o...}'. Henceforth we assume that the input string
is stored as  UTF-8 so that each ASCII character uses a byte.
We load the first 16~bytes (\texttt{<!doctype html><}).
We find the target characters: the first character and the sixteenth~character are a match (\asciicharacter{<}). Algorithm~\ref{alg:vector_classification} returns the location of the first matching character (the first index).
If we must then find the next matching character, we might 
need to start again from the next position, immediately after the first matching character, e.g., we might load the sequence `\texttt{!doctype html><h}'. We then match the same character (\asciicharacter{<}), this time as the fifteenth character. We are duplicating the effort, reprocessing many of the same characters twice.\footnote{In practice, the Web engine might call a new routine to scan tag names without vectorization, which reduces the likelihood of needing to reload and reclassify the same characters.} 

Instead, we can adopt an approach similar to that used by systems like the simdjson library~\cite{langdale2019parsing}.
We load non-overlapping blocks of 64~characters. Each block of 64~characters is turned into a 64-bit register where each bit in the register corresponds to a loaded character. If the character is a match, then the corresponding bit is set to 1. The computed 64-bit word serves as an index for the 64~characters. We can iterate over the set bits of the 64-bit word, as each corresponds to a matching character. Once we have used up this index, we can load another block of 64~characters and so forth. In this manner, we do not duplicate the effort: we load each block of 64~characters only once to match characters. See \S~\ref{sec:sixtyfou} for details.

\section{Instruction Sets}

There is a wide range of processors
supporting vector or SIMD instructions (POWER, RISC-V, etc.). However, two types of
commodity processors dominate: x64 and ARM.

Most common on PCs and servers are x64 processors generally made by Intel and AMD\@.
These processors fall into many microarchitecture categories, including the following levels~\cite{microlevels}:
\begin{itemize}
\item \texttt{x86-64-v1}: SSE2 (16-byte registers);
\item \texttt{x86-64-v2}:  SSE3, SSSE3, SSE~4.1, SSE~4.2, POPCNT;
\item \texttt{x86-64-v3}:  AVX, AVX2 (32-byte registers), BMI1, BMI2, LZCNT;
\item \texttt{x86-64-v4}:  AVX-512 (AVX512F, AVX512BW, AVX512CD, AVX512DQ, AVX512VL)  (64-byte registers). 
\end{itemize}
The features are additive: a processor of level \texttt{x86-64-v2} has the features of level \texttt{x86-64-v1}, and so forth.
From a historical perspective, \texttt{x86-64-v1} goes back to 2003 (AMD K8), \texttt{x86-64-v2} to 2008 (Intel Nehalem), \texttt{x86-64-v3} to 2013 (Intel Haswell) and \texttt{x86-64-v4} to 2017 (Intel Skylake-X).
Recent releases of the Red Hat and SUSE Linux systems are built for systems supporting \texttt{x86-64-v2}.
Recent releases of Windows~11 (version 24H2 and following) require SSE~4.2 which matches level \texttt{x86-64-v2}.
More recent processors from Intel and AMD support even more advanced SIMD instructions, but there is not yet an agreement on the \texttt{x86-64-v5} level at this time.
When compiling software for an x64 processor, the simplest approach might be to pick one of the lowest levels as a target (e.g., \texttt{x86-64-v1}). It is also common to determine the features of the CPU at runtime, but it requires more care.

Most mobile devices (e.g., smartphones, tablets) and some servers and laptops run on 64-bit ARM processors (aarch64). All of these processors share the same SIMD instructions with 16-byte registers: ARM NEON also called Advanced SIMD\@. There are a few extensions: e.g., for dot products and 16-bit floating-point numbers. More recent processors also support other SIMD instruction sets: SVE and SVE2~\cite{stephens2017arm}. However, relatively few ARM systems support SVE or SVE2 at this time.

ARM processors only offered a 64-bit mode starting with ARMv8-A architecture in 2011. It was first broadly available in the iPhone~5s, launched in 2013.
Thus, whereas x64 systems have different levels of SIMD instruction sets,  the 64-bit ARM processors are more uniform from an instruction-set perspective. In part due to their shorter history, 64-bit ARM processors have a more predictable set of instructions: they all support ARM NEON with little variation.

The WebKit engine is developed by Apple. In turn, current Apple  systems use 64-bit ARM processors, whether they are laptops, phones, or tablets. 
The Blink engine is developed by Google. It is designed to build on a wide range of platforms, including Android devices and 
various versions of Windows. Most Android devices have ARM processors.
As of July~2024, the vectorized classification is only enabled under ARM~platforms in Blink.

The algorithms and techniques that we consider are not specific to ARM~NEON. Yet, for simplicity, we focus on ARM~NEON. In our context, it is the most relevant instruction set.

\subsection{ARM Instructions}

A detailed discussion of the 64-bit ARM instruction set is beyond our scope. However, it is useful to review a relevant subset.
Compilers often produce regular instructions such as the following:
\begin{itemize}
    \item \texttt{ldr}: loads a value from memory into a register.
\item \texttt{cbnz}: compares a register with zero and branches if it is not zero.
\item \texttt{rbit}: reverses the bit order in a register.
\item \texttt{clz}: counts the number of leading zero bits in a register.
\end{itemize}

Programming languages such as C or C++ do not allow the programmer to directly express SIMD operations. To avoid having to write assembly, we often use intrinsic functions. Intrinsic functions are recognized by the compiler as representing low-level operations, often corresponding directly to CPU instructions. They are not part of the standard library but are provided by the compiler or the vendor. There are many intrinsic functions and many SIMD instructions. The following instructions are relevant to our problems:

\begin{itemize}
   \item  \texttt{ld1}: loads elements from memory into a vector register.
To emit this instruction, we can use intrinsic functions such as \texttt{vld1q\_u8} for unsigned 8-bit integers. There are other similar loads and store instructions (\texttt{ld2}, \texttt{st1}, \texttt{st2}, \ldots).
   
   \item \texttt{cmeq}: compares two vectors for equality, it compares each pair of values and sets the result to all ones (\texttt{0xF\ldots F}) if they are equal, otherwise to zero. We can use the intrinsic function \texttt{vceqq\_u8} to compare unsigned 8-bit integers.

   \item \texttt{and}: computes the bitwise and between two vectors. E.g., we may use the \texttt{vandq\_u8} intrinsic.

   \item \texttt{orn}: performs a bitwise OR operation between a register and the bitwise NOT of another register. E.g., we may use the \texttt{vornq\_u8} intrinsic.

   \item \texttt{add}: adds the components of two vectors (\texttt{vaddq\_u8} intrinsic).
   
   \item \texttt{umaxv}: finds the maximum value in a vector.
   The intrinsic \texttt{vmaxvq\_u32} finds the maximum of unsigned 32-bit integers and the intrinsic \texttt{vmaxvq\_u8} finds the maximum of unsigned 8-bit integers.
   
      \item \texttt{uminv}: finds the minimum value in a vector. It has intrinsic functions similar to \texttt{umaxv}: \texttt{vminvq\_u32} for unsigned 32-bit integers, \texttt{vminvq\_u16} for unsigned 16-bit integers, \texttt{vminvq\_u8} for unsigned 8-bit integers.

   \item \texttt{shrn}: shifts each element in a vector to the right and narrows the result. For example, each 16-bit word might be shifted right and we keep only the least significant 8~bits. A 16-byte inputs results in an 8-byte output. We might use the \texttt{vshrn\_n\_u16} intrinsic for unsigned 16-bit integers.

   \item \texttt{umov}: moves a value from a vector to a general-purpose register, it can be used in conjunction with the instructions  \texttt{shrn}, \texttt{umaxv} and \texttt{uminv}. We have several intrinsic functions depending on the desired bit widths such 
   as \texttt{vgetq\_lane\_u64},   \texttt{vgetq\_lane\_u32},  \texttt{vgetq\_lane\_u16},  \texttt{vgetq\_lane\_u8}. Some compilers 
   generate the \texttt{fmov} instruction (floating-point move) instead of the \texttt{umov} instruction.  
   
   \item \texttt{addp}: adds pairs of adjacent elements in a vector. E.g., adds pairs of 8-bit integers and produce a 16-bit output. Thus a vector of sixteen 8-bit values becomes a vector of eight 16-bit values. We may use intrinsic functions such as \texttt{vpaddq\_u8} for unsigned 8-bit integers.

   \item \texttt{tbl}: performs a table lookup on a vector. It uses elements from one vector to index into another vector, effectively performing a gather operation. 
    If the index value is within the bounds of the table vectors (e.g., $0,1,\ldots, 15$), the corresponding byte from the table is copied to the result vector. If the index is out of bounds, a zero is placed in the result for that element.
   We may use the intrinsic function \texttt{vtbl1q\_u8} to generate the \texttt{tbl} instruction.
\end{itemize}

Table~\ref{tab:latinst} presents the latency and throughput of these SIMD instructions on two ARM microarchitectures: Apple Avalanche~\cite{applesilicon} and Neoverse~V2~\cite{neoversev2}. 
To our knowledge, Apple does not release official latencies and throughputs for its processors, so these values are estimates.
The Apple Avalanche is likely representative of Apple's performance cores. We believe that the Neoverse~V2 is the basis for Amazon's Graviton~4 server processor.
The Apple microarchitecture has high throughput: most instructions can be issued four times per cycle while only simple instructions such as \texttt{cmeq}, \texttt{and}, \texttt{orn}, \texttt{add} and \texttt{addp} have a comparable throughput on Neoverse~V2. 
The \texttt{uminv} and \texttt{umaxv} instructions on the Apple microarchitecture have the same performance irrespective of the bit width (8-bit, 16-bit, 32-bit) but on the Neoverse~V2, the 8-bit version is relatively slow: one instruction of throughput and 4~cycles of latency. 
There are many more ARM-based microarchitectures and a complete survey is outside our scope. 
\begin{table}
    \centering
\begin{tabular}{lcccc}
\toprule
                                       & \multicolumn{2}{c}{Neoverse~V2}  & \multicolumn{2}{c}{Apple Firestorm}\\
                                       & L & T & L & T  \\\midrule
                                       
\texttt{cmeq}                          & 2 & 4        & 2         & 4\\
\texttt{and}                          & 2 & 4        & 2         & 4\\
\texttt{orn}                          & 2 & 4        & 2         & 4\\
\texttt{add}                          & 2 & 4        & 2         & 4\\
\texttt{uminv}/\texttt{umaxv} (32-bit) & 2 & 2        & 3         & 4 \\
\texttt{uminv}/\texttt{umaxv} (16-bit) & 4 & 2        & 3         & 4 \\
\texttt{uminv}/\texttt{umaxv} (8-bit)  & 4 & 1        & 3         & 4 \\
\texttt{umov}                          & 2 & 1        & $\leq 10$ & 2 \\
\texttt{shrn}                          & 2 & 2        & 3         & 4 \\
\texttt{addp}                          & 2 & 4        & 2         & 4 \\
\texttt{tbl}                           & 2 & 2        & 2         & 4 \\\bottomrule\restartrowcolors{}
    \end{tabular}
    \caption{Latency and throughput of various SIMD instructions for two microarchitecture. The latency is the time in cycles required to complete the instruction. The throughput is the number of instructions that can be issued per cycle.}
    \label{tab:latinst}
\end{table}

\section{Efficient SIMD Implementation}

We can use SIMD instructions deliberately from a language like C++ in various ways: inline assembly, intrinsic functions, etc. Both WebKit and Blink use dedicated SIMD libraries. These libraries abstract away 
the instruction sets, allowing the engineers to focus on the programming
logic. Furthermore, they allow the engineers to quickly support a wider
range of platforms with less programming effort. 
We describe the WebKit and Blink implementations as of July~2024. The source code is freely available
online.\footnote{\url{https://github.com/WebKit/WebKit} and \url{https://github.com/chromium/chromium/tree/main/third_party/blink}}

\subsection{WebKit Implementation}
\label{sec:webkit}
We present in Fig.~\ref{fig:simplewebkit} a slightly simplified version of the WebKit implementation. The full version also supports UTF-16 inputs, but most Web sites use UTF-8. Further, we also
apply the simplifying assumption that the input contains at least 16~bytes.
WebKit relies on  SIMDe~\cite{simde}. The SIMDe \texttt{simde\_uint8x16\_t} type represents
a vector of 16~eight-bit unsigned integers.

The \texttt{scanText} function is part of WebKit's HTML parser. It scans a buffer of text for specific characters and extracts a substring based on those findings. 
 If a null terminator is found within the buffer or if the first character encountered is \asciicharacter{\&} or
\asciicharacter{\textbackslash{}r}, it signals a parsing error or the need for special handling and returns an empty string.
 The \texttt{vectorEquals8Bit} function
does the vectorized classification, returning is the SIMD equivalent.\footnote{The original WebKit implementation was based on comparisons while the original Blink was based on  vectorized classification: the WebKit implementation adopted vectorized classification later.}
The \texttt{vectorMatch} lambda function uses the \texttt{vectorEquals8Bit} function to find the first occurrence of the target characters within a SIMD vector.  

The \texttt{vectorEquals8Bit} function identifies the matching character with as few as three instructions: a bitwise AND (\texttt{and} instruction), a vectorized table lookup (\texttt{tbl} instruction), and a byte-wise comparison (\texttt{cmeq} instruction). The algorithm is based on the observation that the four characters that we need to identify (\asciicharacter{<}, \asciicharacter{\textbackslash{}r}, \asciicharacter{\&}, \asciicharacter{\textbackslash{}0}) have code point values that are \texttt{0x00}, \texttt{0x26}, \texttt{0x3c}, and \texttt{0x0d}: they can be distinguished by the least significant 4~bits (\texttt{0x0}, \texttt{0x6}, \texttt{0xc}, and \texttt{0xd}). Hence the algorithm proceeds as follows:
\begin{enumerate}
    \item Take the character value $c$ as an 8-bit integer.
    \item Use a 16-element table $t$ where all values are zero except for the those corresponding to indexes \texttt{0x0}, \texttt{0x6}, \texttt{0xc}, and \texttt{0xd} where we store \texttt{0x00}, \texttt{0x26}, \texttt{0x3c}, and \texttt{0x0d}.
    \item Use the least significant bits of $c$, computed as $c \mathrm{~AND~0xF}$ to look up a value in the table. We have that $t[c \mathrm{~AND~0xF}] = c$ if and only if $c$ is one of \texttt{0x00}, \texttt{0x26}, \texttt{0x3c}, and \texttt{0x0d}.
    
\end{enumerate}
Because it is implemented using SIMD instructions, sixteen bytes are processed at once.
Thus, using as few as three instructions, we get to a SIMD register where the matching characters correspond to a true value (\texttt{0xFF} by convention).

Given this SIMD register, we need to find the index of the first match, if any.
We provide the implementation in Fig.~\ref{fig:simplewebkitfindfirst}.
The WebKit engineers first compute the maximum byte value in the register using a function from the SIMDe~\cite{simde} framework (\texttt{simde\_vmaxvq\_u8} corresponding to the \texttt{vmaxvq} instruction): this function returns 0 if and only if there is no match. Otherwise, the value \texttt{0xFF} is returned. In this case,
the function continues: it computes a byte-wise
OR NOT (using the \texttt{orn} instruction) between the values 0, 1, \ldots 15 and our SIMD register. It then calls \texttt{simde\_vminvq\_u8} (corresponding to the \texttt{vminvq} instruction). These two instructions (\texttt{orn} and \texttt{vminvq}) are sufficient to return the smallest index. E.g., suppose that only the second and last character were a match, then we have the SIMD register (0, \texttt{0xFF}, 0, 0, 0, 0, 0, 0, 0, 0, 0, 0, 0, 0, 0, \texttt{0xFF}). After the bitwise OR NOT with the values  0, 1, \ldots 15, we get  (\texttt{0xFF}, 1, \texttt{0xFF}, \texttt{0xFF}, \texttt{0xFF}, \texttt{0xFF}, \texttt{0xFF}, \texttt{0xFF}, \texttt{0xFF}, \texttt{0xFF}, \texttt{0xFF}, \texttt{0xFF}, \texttt{0xFF}, \texttt{0xFF}, \texttt{0xFF}, 15), and the minimum is 1, indicating the first match.

\begin{figure}
    \begin{tabular}{l}
\begin{lstlisting}[style=customcpp]
String scanText() {
  auto* start = m_parsingBuffer.position();
  auto* end = start + m_parsingBuffer.lengthRemaining();
  auto vectorEquals8Bit = [&](auto input) {
    simde_uint8x16_t lowNibbleMask { '\0', 0, 0, 0, 0, 0, '&', 0, 0, 0, 0, 0, '<', '\r', 0, 0 };
    simde_uint8x16_t v0f = SIMD::splat<uint8_t>(0x0f);
    return SIMD::equal(simde_vqtbl1q_u8(lowNibbleMask, SIMD::bitAnd(input, v0f)), input);
  };
  CharacterType* cursor = nullptr;
  auto vectorMatch = [&](auto input) {
    return SIMD::findFirstNonZeroIndex(vectorEquals8Bit(input));
  };
  cursor = SIMD::find(std::span { start, end }, vectorMatch);
  m_parsingBuffer.setPosition(cursor);
  if (cursor != end) {
    if (*cursor == '\0')
      return didFail(HTMLFastPathResult::FailedContainsNull, String());

    if (*cursor == '&' || *cursor == '\r') {
      m_parsingBuffer.setPosition(start);
      return scanEscapedText();
    }
  }
  unsigned length = cursor - start;
  return length ? String({ start, length }) : String();
}
\end{lstlisting}
    \end{tabular}
    \caption{WebKit Implementation}
    \label{fig:simplewebkit}
\end{figure}

\begin{figure}
    \begin{tabular}{l}
\begin{lstlisting}[style=customcpp]
std::optional<uint8_t> findFirstNonZeroIndex(simde_uint8x16_t value) {
    if (!simde_vmaxvq_u8(value))
        return std::nullopt;
    constexpr simde_uint8x16_t indexMask { 0, 1, 2, 3, 4, 5, 6, 7, 8, 9, 10, 11, 12, 13, 14, 15 };
    return simde_vminvq_u8(simde_vornq_u8(indexMask, value));
}
\end{lstlisting}
    \end{tabular}
    \caption{WebKit's Find-First Implementation}
    \label{fig:simplewebkitfindfirst}
\end{figure}

\begin{figure}
    \begin{tabular}{l}
\begin{lstlisting}[style=customcpp]
const char* find(std::span<const char> span, const auto& vectorMatch) {
    auto* cursor = span.data();
    auto* end = span.data() + span.size();
    for (; cursor + (16 - 1) < end; cursor += 16) {
            if (auto index = vectorMatch(SIMD::load(cursor)))
                return cursor + index.value();
    }
    if (cursor < end) {
        if (auto index = vectorMatch(SIMD::load(end - stride)))
            return end - stride + index.value();
    }
    return end;
}
\end{lstlisting}
    \end{tabular}
    \caption{WebKit's Find Implementation}
    \label{fig:simplewebkitfind}
\end{figure}

\begin{figure}
    \begin{tabular}{l}
\begin{lstlisting}[style=customasmarm]
        and     v3.16b, v2.16b, v0.16b; AND 0xF
        tbl     v3.16b, { v1.16b }, v3.16b ; table lookup
        cmeq    v2.16b, v3.16b, v2.16b ; compare 
        umaxv   b3, v2.16b ; check if the result is zero
        fmov    w9, s3 ; could be umov
        cbnz    w9, MATCH 
; ... insert code to loop back
MATCH: ; we have a match
        orn     v0.16b, v0.16b, v2.16b
        uminv   b0, v0.16b
        fmov    w8, s0 ; could be umov
        add     x0, x0, x8 ; x0 contains the index of a match

\end{lstlisting}
    \end{tabular}
    \caption{WebKit's Possible ARM NEON Assembly}
    \label{fig:simplewebkitasm}
\end{figure}
The  function \texttt{SIMD::find} searches the buffer for a match using the \texttt{vectorMatch} function. See Fig.~\ref{fig:simplewebkitfind}.
Under ARM, the core of the find function might compile to the assembly instruction in Fig.~\ref{fig:simplewebkitasm}.
When a 16-byte input does not contain any matching character, we use about seven instructions---and slightly more in practice:  (\texttt{ldr}, \texttt{and}, \texttt{tbl}, \texttt{cmeq}, \texttt{umaxv}, \texttt{cbnz}). Overall, we use fewer than 16~instructions to process 16~bytes. When the 16~bytes contain a match, we must add another series of instructions (\texttt{orn}, \texttt{uminv}, \texttt{umov}/\texttt{fmov}, \texttt{add}), making the process more expensive: we get closer to one instruction per byte.
The \texttt{umov} or \texttt{fmov} instructions move the data from a SIMD register to a general-purpose register while the \texttt{add} instruction increments our index.

The exact performance of the code depends on the processor microarchitecture. However, we should pay attention to the 
\texttt{uminv} and \texttt{umaxv} instructions, which can require many cycles of latency (e.g.,~four) on some processors.

\subsection{Blink Implementation}
\label{sec:blink}

The Blink implementation resembles the WebKit implementation. Instead of SIMDe, it relies on the Highway library~\cite{highway}. The
Highway library provides a \texttt{FindFirstTrue}~function which is equivalent in functionality to WebKit's \texttt{findFirstNonZeroIndex} function (see Fig~\ref{fig:simplewebkitfindfirst}).
Unlike the WebKit implementation, the \texttt{FindFirstTrue}~function in Highway does not use the \texttt{umaxv} and \texttt{uminv} instructions. Instead, Highway uses the
\texttt{vshrn\_n\_u16} intrinsic function. This intrinsic function corresponds to the \texttt{shrn} instruction. It takes each 16-bit word, shifts them to the right by 4~bits, and then keeps only the least significant 8~bits out of each 16-bit word. The net result covers 64~bits and can be moved to a general-purpose register. To see why it is
a helpful operation,  consider again the mask where only the first and last character were a match ( \texttt{0xFF}, 0, 0, 0, 0, 0, 0, 0, 0, 0, 0, 0, 0, 0, 0, \texttt{0xFF}).
\begin{enumerate}
    \item We first represent it as eight 16-bit words with the little-endian convention:
\texttt{0x00FF}, \texttt{0x0000}, \texttt{0x0000}, \texttt{0x0000}, \texttt{0x0000}, \texttt{0x0000}, \texttt{0x0000} , \texttt{0xFF00}.
\item We shift by 4~bits:
\texttt{0x000F}, \texttt{0x0000}, \texttt{0x0000}, \texttt{0x0000}, \texttt{0x0000}, \texttt{0x0000}, \texttt{0x0000} , \texttt{0x0FF0}.
\item We keep only the least significant byte of each 16-bit word:
\texttt{0x0F}, \texttt{0x00}, \texttt{0x00}, \texttt{0x00}, \texttt{0x00}, \texttt{0x00}, \texttt{0x00} , \texttt{0xF0}.
\item We convert the result to a 64-bit word: \texttt{0xF0000000000000F0}.
\end{enumerate}
The final 64-bit still contains the information concerning the matches: each matched character corresponds to a 4-bit subword. To find the index of the first match, it suffices to compute the number of trailing zeros and divide this number by four.
Fig.~\ref{fig:simplechromiumasm} presents a possible compiled assembly. Observe how computing the number of trailing zeros is done using two instructions: \texttt{rbit} which reverses the bits, and  \texttt{clz} which computes the number of leading zeros.\footnote{ARM added new relevant functionality to its designs as an extension, called \emph{Common Short Sequence Compression} (\texttt{FEAT\_CSSC}). This extension adds a trailing-zero instruction as well as a population count function, among others. We expect that it will be supported by upcoming processors.}

\begin{figure}
    \begin{tabular}{l}
\begin{lstlisting}[style=customasmarm]
        and     v30.16b, v29.16b, v30.16b
        tbl     v31.16b, {v31.16b}, v30.16b
        cmeq    v31.16b, v31.16b, v29.16b
        shrn    v31.8b, v31.8h, 4
        umov    x0, v31.d[0]
        cbz     x0, .NOMATCH ; does not match
        rbit    x0, x0
        clz     x0, x0
        lsr     x0, x0, 2; division by 4
\end{lstlisting}
    \end{tabular}
    \caption{Blink's Possible ARM NEON Assembly}
    \label{fig:simplechromiumasm}
\end{figure}

\subsection{64-bit Vectorized Classification}
\label{sec:sixtyfou}
Both the WebKit and Blink implementations do some unnecessary processing, potentially reloading and reclassifying the same characters many times. Further, they might be limited by the latency of the instructions used to identify the first matching character in a block of 16~bytes.

Instead, let us load non-overlapping blocks of 64~bytes and identify the matching characters in a 64-bit word, as in  Fig.~\ref{fig:fff}. Given this 64-bit index, 
we can check whether it is zero, and if so, we know that we must advance forward by 64~bytes in the input string. Otherwise, we just compute the number of trailing zeros and advance the string by the count. We can shift the index by the corresponding number of trailing zeros plus one, and if the result is non-zero then the index can be used to advance in the string once more, and so forth. If the result is zero, we need to load 64~new~bytes: we always load the bytes in a non-overlapping manner, to the next 64~bytes. At the end of the string, when fewer than 64~bytes remain, we may need to either fall back on a conventional routine.

\begin{figure}
   \centering
    \begin{tabular}{l}

\tikzstyle{block} = [rectangle, 
 draw, fill=blue!20, 
    text width=10em, text centered, rounded corners, minimum height=2em]
\tikzstyle{line} = [draw, -latex']
\tikzset{node distance=5cm and 0.1cm}
\scalebox{0.7}{
\begin{tikzpicture}[auto]

    \node [block] (load1) {vld1q\_u8(buffer)};
    \node [block, right of=load1] (load2) {vld1q\_u8(buffer+16)};
    \node [block, right of=load2] (load3) {vld1q\_u8(buffer+32)};
    \node [block, right of=load3] (load4) {vld1q\_u8(buffer+48)};

    \node [block, below = 1cm of load1] (low1) {vqtbl1q\_u8};
    \node [block, right of=low1] (low2) {vqtbl1q\_u8};
    \node [block, right of=low2] (low3) {vqtbl1q\_u8};
    \node [block, right of=low3] (low4) {vqtbl1q\_u8};

    \node [block, below = 1cm of low1] (cmp1) {vceqq\_u8};
    \node [block, right of=cmp1] (cmp2) {vceqq\_u8};
    \node [block, right of=cmp2] (cmp3) {vceqq\_u8};
    \node [block, right of=cmp3] (cmp4) {vceqq\_u8};

    \node [block, below = 1cm of cmp1] (mask1) {AND bit\_mask};
    \node [block, right of=mask1] (mask2) {AND bit\_mask};
    \node [block, right of=mask2] (mask3) {AND bit\_mask};
    \node [block, right of=mask3] (mask4) {AND bit\_mask};

    \node [block, below = 1cm of mask1] (add1) {vpaddq\_u8};
    \node [block, right of=add1] (add2) {vpaddq\_u8};
    \node [block, below = 1cm of add1] (add3) {vpaddq\_u8};
    \node [block, below = 1cm of add3] (add4) {vpaddq\_u8};

    \path [line] (load1) -- (low1);
    \path [line] (load2) -- (low2);
    \path [line] (load3) -- (low3);
    \path [line] (load4) -- (low4);

    \path [line] (low1) -- (cmp1);
    \path [line] (low2) -- (cmp2);
    \path [line] (low3) -- (cmp3);
    \path [line] (low4) -- (cmp4);

    \path [line] (cmp1) -- (mask1);
    \path [line] (cmp2) -- (mask2);
    \path [line] (cmp3) -- (mask3);
    \path [line] (cmp4) -- (mask4);

    \path [line] (mask1) -- (add1);
    \path [line] (mask2) -- (add1);
    \path [line] (mask3) -- (add2);
    \path [line] (mask4) -- (add2);

    \path [line] (add1) -- (add3);
    \path [line] (add2) -- (add3);

    \path [line] (add3) -- (add4);

    \node [below = 1cm of add4] (result) {sum};
    \path [line] (add4) -- (result);

\end{tikzpicture}}
    \end{tabular}
    \caption{Computation of a 64-bit index: observe the obvious parallelism }
    \label{fig:dia}
\end{figure}
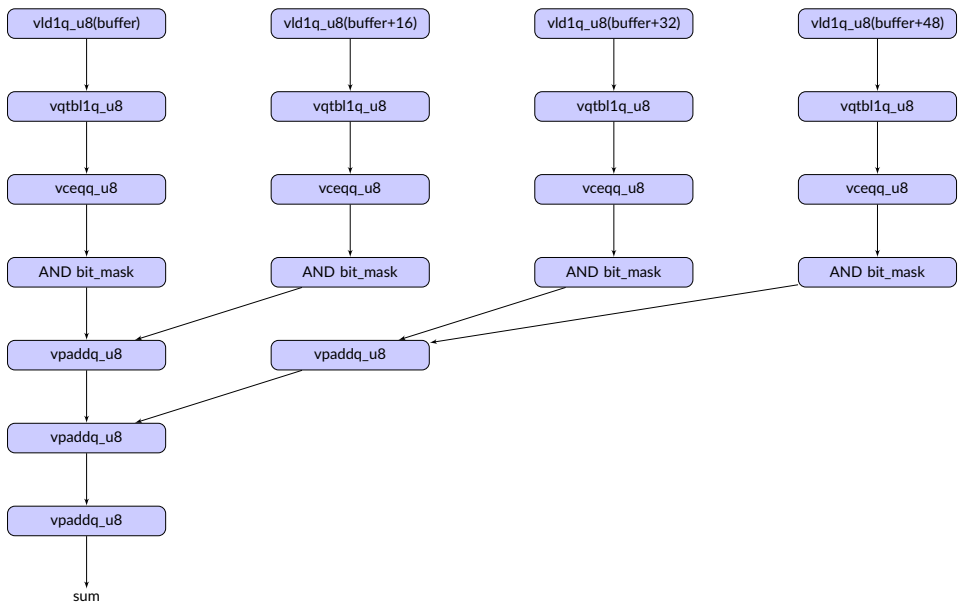

The routine in Fig.~\ref{fig:fff} begins by filling four 16-byte registers corresponding to a block of 64~bytes using the \texttt{vld1q\_u8} intrinsic---corresponding to the \texttt{ld1} instructions. We then use the same vectorized classification as in the WebKit and Blink routines: we
select the least significant 4~bits of each byte with a bitwise AND (\texttt{\&} operator) followed by a vectorized table lookup (\texttt{vqtbl1q\_u8} intrinsic, \texttt{tbl} instructions) followed by a comparison (\texttt{vceqq\_u8} intrinsic, \texttt{cmeq} instruction).
We then have 64~bytes where the matching characters have
been identified and stored in \texttt{matchesones1}, \texttt{matchesones2}, \texttt{matchesones3}, and \texttt{matchesones4}.
If the microarchitecture allows it, these computations can be executed in parallel: many ARM processors have 4~SIMD execution units.
Finally, 
we must aggregate these four 16-byte registers counting byte values \texttt{0x00} and \texttt{0xFF} into a 64-bit integer. See Fig.~\ref{fig:dia} for an illustration.

\begin{enumerate}
    \item We compute the bitwise AND of the four SIMD variables with 
    (\texttt{0x01}, \texttt{0x02}, \texttt{0x4}, \texttt{0x8}, \texttt{0x10}, \texttt{0x20}, \texttt{0x40}, \texttt{0x80},
                                 \texttt{0x01}, \texttt{0x02}, \texttt{0x4}, \texttt{0x8}, \texttt{0x10}, \texttt{0x20}, \texttt{0x40}, \texttt{0x80}). 
These values correspond to bits at position 1, 2, 3, \ldots 8, repeated twice.
\item We do horizontal pairwise byte sums (\texttt{vpaddq\_u8} intrinsic, \texttt{addp} instruction): starting with two 16-byte values, treated as a sequence of 32~bytes, we sum the first two bytes and write the result in the output, we sum the next two bytes and write the result as a byte and so forth. By doing it four times, we collapse the initially 64~bytes into the desired 64-bit integer.
To illustrate the algorithm, suppose that all of the characters matched. Then we would do a horizontal pairwise byte sum on the 32-byte sequence
 (\texttt{0x01}, \texttt{0x02}, \ldots, \texttt{0x80}). We want to show that we can compute the 64-bit word \texttt{0xFFFFFFFFFFFFFFFF}.
 Initially, we get the 16-byte sequence
  (\texttt{0x03}, \texttt{0x0c}, \texttt{0x30}, \texttt{0xc0}, 
  \texttt{0x03}, \texttt{0x0c}, \texttt{0x30}, \texttt{0xc0},
  \texttt{0x03}, \texttt{0x0c}, \texttt{0x30}, \texttt{0xc0},
  \texttt{0x03}, \texttt{0x0c}, \texttt{0x30}, \texttt{0xc0}).
      We get two such sequences (\texttt{sum0} and \texttt{sum1} in Fig.~\ref{fig:fff}). 
      Combining the once more
      we get the 16-byte sequence 
      \texttt{0x0F}, \texttt{0xF0}, \texttt{0x0F}, \texttt{0xF0}, 
            \texttt{0x0F}, \texttt{0xF0}, \texttt{0x0F}, \texttt{0xF0}, 
                  \texttt{0x0F}, \texttt{0xF0}, \texttt{0x0F}, \texttt{0xF0}, 
                        \texttt{0x0F}, \texttt{0xF0}, \texttt{0x0F}, \texttt{0xF0}.
Repeating one last time, this time by repeating the same sequence of 16-byte values twice, we finally get the 64-bit value \texttt{0xFFFFFFFFFFFFFFFF} twice, as expected. We keep only one of the two 64-bit values. If some characters were unmatched, then zero bits would appear.
Depending on the microarchitecture, the first two \texttt{addp} instructions can be executed simultaneously, whereas the next two must be done in sequence. Though there is likely some non-trivial latency, it is amortized over an input of 64~bytes of input data.
\end{enumerate}
Because we need to compute the number of trailing zeros if the output is non-zero, we might reverse the bits (\texttt{rbit}) once. In this manner, we do not need to repeatedly call the sequence \texttt{rbit}/\texttt{clz} to compute the number of trailing zeros, if there are several matches: it suffices to call \texttt{rbit} once followed by a single call to \texttt{clz}. However, this may not be a significant optimization in practice.


\begin{figure}
    \begin{tabular}{l}
\begin{lstlisting}[style=customcpp]
// constants:
static uint8x16_t low_nibble_mask = {0, 0, 0, 0, 0, 0, 0x26, 0, 0, 0, 0, 0, 0x3c, 0xd, 0, 0};
static uint8x16_t bit_mask = {0x01, 0x02, 0x4, 0x8, 0x10, 0x20, 0x40, 0x80,
                                 0x01, 0x02, 0x4, 0x8, 0x10, 0x20, 0x40, 0x80};
static uint8x16_t v0f = vmovq_n_u8(0xf);

uint64_t update(const uint8_t *buffer) {
    uint8x16_t data1 = vld1q_u8(buffer);
    uint8x16_t data2 = vld1q_u8(buffer + 16);
    uint8x16_t data3 = vld1q_u8(buffer + 32);
    uint8x16_t data4 = vld1q_u8(buffer + 48);
    uint8x16_t lowpart1 = vqtbl1q_u8(low_nibble_mask, data1 & v0f);
    uint8x16_t lowpart2 = vqtbl1q_u8(low_nibble_mask, data2 & v0f);
    uint8x16_t lowpart3 = vqtbl1q_u8(low_nibble_mask, data3 & v0f);
    uint8x16_t lowpart4 = vqtbl1q_u8(low_nibble_mask, data4 & v0f);
    uint8x16_t matchesones1 = vceqq_u8(lowpart1, data1);
    uint8x16_t matchesones2 = vceqq_u8(lowpart2, data2);
    uint8x16_t matchesones3 = vceqq_u8(lowpart3, data3);
    uint8x16_t matchesones4 = vceqq_u8(lowpart4, data4);
    uint8x16_t sum0 = vpaddq_u8(matchesones1 & bit_mask, matchesones2 & bit_mask);
    uint8x16_t sum1 = vpaddq_u8(matchesones3 & bit_mask, matchesones4 & bit_mask);
    sum0 = vpaddq_u8(sum0, sum1);
    sum0 = vpaddq_u8(sum0, sum0);
    return vgetq_lane_u64(vreinterpretq_u64_u8(sum0), 0);
}
\end{lstlisting}
    \end{tabular}
    \caption{Computation of a 64-bit index over 64~bytes using ARM NEON intrinsic functions}
    \label{fig:fff}
\end{figure}

\section{Experiments}

We implemented the WebKit and Blink approaches, as well as the 64-bit indexing approach. As a reference implementation, we also use the standard C++ function (\texttt{string::find\_first\_of}) as provided by the runtime library. We make our code freely available to ease reproducibility.\footnote{\url{https://github.com/lemire/htmlscanning}}
We use two different ARM-based systems for benchmarking. See Table~\ref{tab:test-cpus}. 

We collected three HTML files by saving the HTML pages of the BBC, Microsoft Office, and Google Web sites. See Table~\ref{tab:test-datasets-size}. Importantly, the BBC Web site we captured has fewer matching characters, relatively 
speaking, than the Google Web site: we should thus expect
that scanning the BBC data should be more expensive on a per input byte basis.
We make the data available with our software.

\begin{table}
\caption{\label{tab:test-cpus} Systems 
}
\centering
\begin{tabular}{ccccccc}\toprule
Processor   &  Frequency  & Microarchitecture                           & Memory  & Compiler\\ \midrule
Graviton 4   & \SI{2.8}{\GHz}  & Neoverse V2 (aarch64, 2022) &  DDR5 (5600\,MT/s)  & GCC  13.2 & \\
Apple M2  & \SI{3.0}{\GHz}  & Avalanche (aarch64, 2022) &  LPDDR5 (6400\,MT/s)  & Apple/LLVM  15 &\\ 
\bottomrule\restartrowcolors{}
\end{tabular}
\end{table}

\begin{table}
\caption{\label{tab:test-datasets-size} Test file sizes and number of matches.}
\centering
\begin{tabular}{l|rr|r}\toprule
file & bytes  & matches & ratio \\ \midrule
BBC & \num{418417} &\num{4420}  & 1.06\% \\
Office & \num{213748} & \num{2393} & 1.12\% \\
Google & \num{20319} & \num{380} & 1.87\%\\\bottomrule
\end{tabular}
\end{table}

\begin{figure}\centering
 \begin{subfigure}[h]{0.49\textwidth}
 \includegraphics[width=0.99\textwidth]{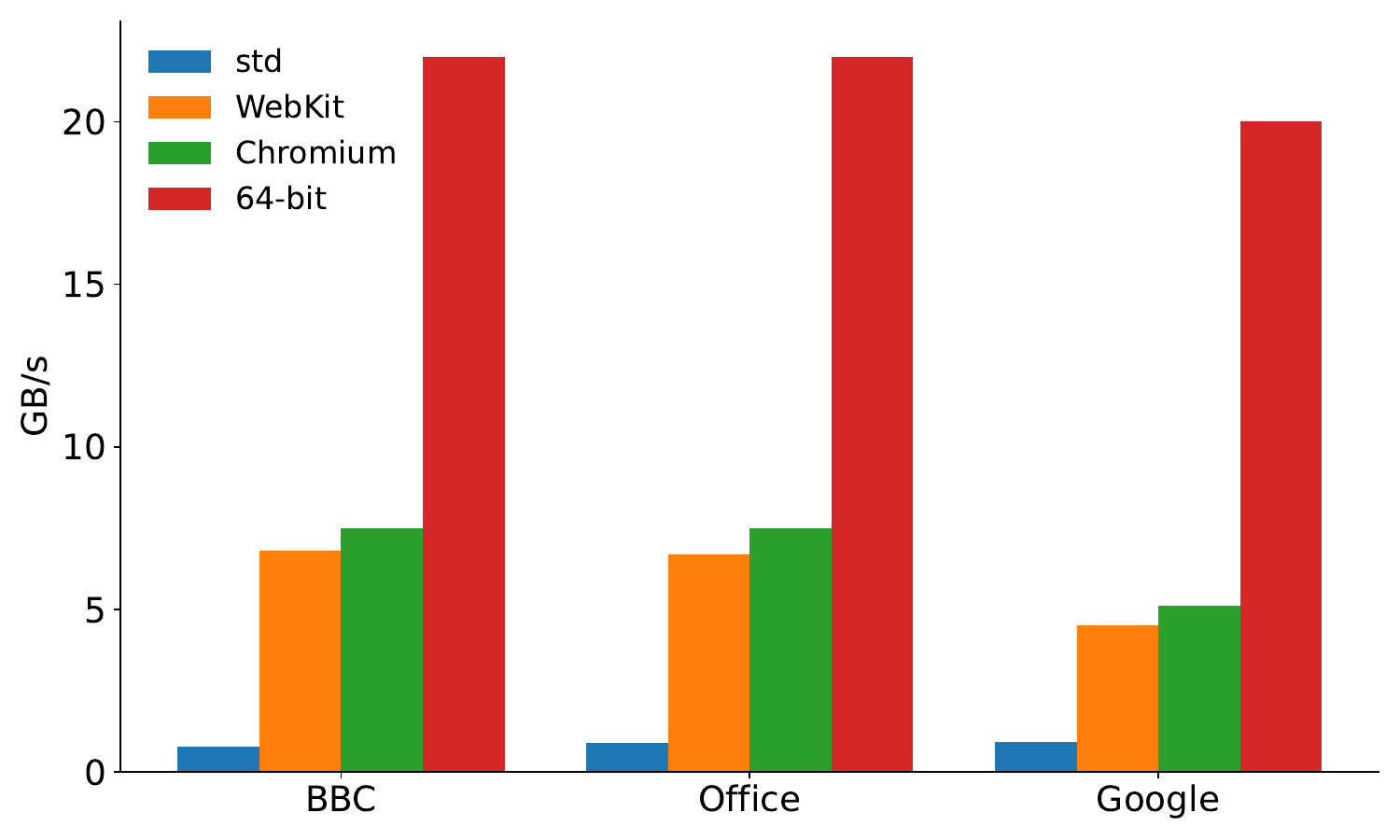}
\caption{Graviton~4} \end{subfigure}
 \begin{subfigure}[h]{0.49\textwidth}
 \includegraphics[width=0.99\textwidth]{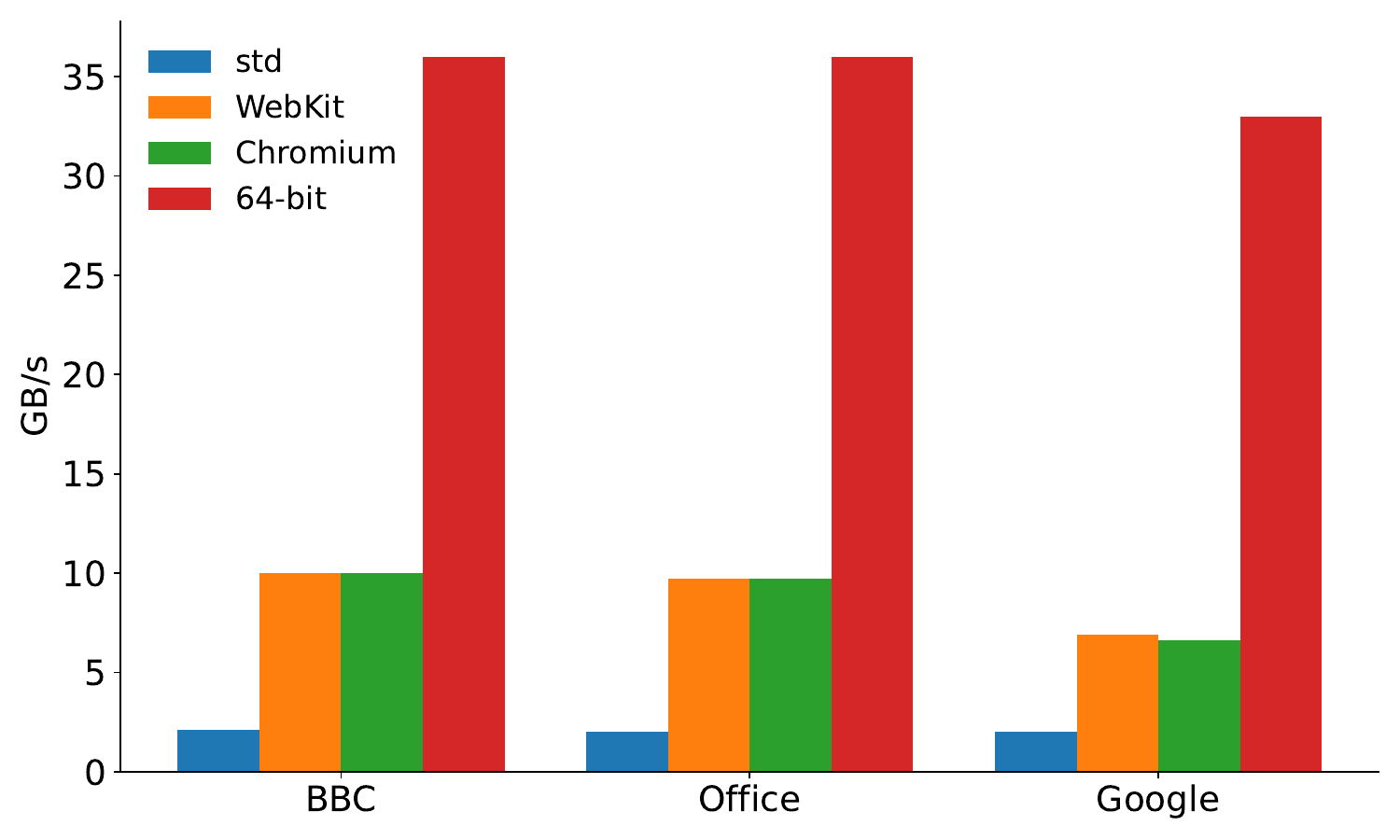}
\caption{Apple~M2} \end{subfigure}
\caption{Parsing speed in gigabytes per second\label{fig:speed}}
\end{figure}

To measure the performance, we take the input
string and we iterate over the matching characters.
We store the matching characters to a volatile variable
to prevent undue optimization. In a Web browser, the scanning would be interleaved with other
operations---but we wish to measure just the scanning.
To help understand our performance, we use performance counters to record the number of instructions retired by cycle as well as the number of instructions per cycle and the effective frequency. 
We present our results in Table~\ref{tab:results} and Fig.~\ref{fig:speed}. We estimate the error of the measured speed to be less than 2\% and the reported number of retired instructions per byte is much less than 1\%.

We find that on the Graviton~4 system, the Blink approach 
is faster than the WebKit approach even though they both 
retire the same number of instructions per byte. The better
performance is due to the higher number of retired instructions for the Blink approach. 

The 64-bit indexing approach is significantly faster than 
the competitors. It is between 15~to 20~times faster than 
the standard function (\texttt{string::find\_first\_of}).
The 64-bit approach is three to four times faster than the WebKit/Blink
approaches. The benefits of the 64-bit approach are slightly greater in the Google file 
which has  more matching characters. This is expected because when several matching characters are within a 64-byte range, we can move from one character to another by merely using the index, without reloading input bytes.

The 64-bit approach requires fewer instructions (40\% less than WebKit/Blink), and it allows us to retire many more instructions per cycle. We save instructions in part by 
amortizing the processing over 64-byte blocks.
In effect, we use a form of batching.

\begin{table}\centering\small
\caption{ \label{tab:results}
Speed in gigabytes per second, number of instructions per cycle and instructions retired per cycle for various data inputs and systems.}
\subfloat[BBC -- Graviton 4]{
\begin{tabular}{lrrr}\toprule
function & GB/s & ins./byte & ins./cycle \\\midrule
std   &  0.78 & 7.8 & 2.16 \\ 
WebKit & 6.8 & 1.0 & 2.51\\
Blink & 7.5 & 1.0 & 2.74\\ 
64-bit & 22 & 0.66 & 5.21\\ 
\bottomrule\restartrowcolors{}
\end{tabular}
}
\subfloat[BBC -- Apple M2]{
\restartrowcolors{}
\begin{tabular}{lrrr}\toprule
function & GB/s & ins./byte & ins./cycle  \\\midrule
std   & 2.1 & 9.0 & 5.42 \\ 
WebKit & 10 & 0.95 & 2.74\\
Blink & 10 & 0.93 & 2.65\\ 
64-bit & 36 & 0.61 & 6.29\\ 
\bottomrule\restartrowcolors{}
\end{tabular}
}\\
\subfloat[Office -- Graviton 4]{
\restartrowcolors{}
\begin{tabular}{lrrr}\toprule
function & GB/s & ins./byte & ins./cycle \\\midrule
std   &  0.90 & 8.7 & 2.82  \\ 
WebKit & 6.7 & 1.0 & 2.46\\
Blink & 7.5 & 1.0 & 2.71\\ 
64-bit & 22 & 0.66 & 5.27\\ 
\bottomrule\restartrowcolors{}
\end{tabular}
}
\subfloat[Office -- Apple M2]{
\restartrowcolors{}
\begin{tabular}{lrrr}\toprule
function & GB/s & ins./byte & ins./cycle  \\\midrule
std   & 2.0 & 9.0 &  5.12 \\ 
WebKit &9.7 & 0.95 &2.72\\
Chromium & 9.7 & 0.93 & 2.57\\ 
64-bit & 36 & 0.62 & 6.44\\ 
\bottomrule\restartrowcolors{}
\end{tabular}
}\\
\subfloat[Google -- Graviton 4]{
\restartrowcolors{}
\begin{tabular}{lrrr}\toprule
function & GB/s & ins./byte & ins./cycle \\\midrule
std   &  0.93 & 7.7 & 2.53 \\ 
WebKit & 4.5 & 1.3 & 2.02\\
Chromium & 5.1 & 1.2 & 2.02\\ 
64-bit & 20 & 0.76 & 5.47\\ 
\bottomrule\restartrowcolors{}
\end{tabular}
}
\subfloat[Google -- Apple M2]{
\restartrowcolors{}
\begin{tabular}{lrrr}\toprule
function & GB/s & ins./byte & ins./cycle  \\\midrule
std   & 2.0 & 9.1 & 5.11 \\ 
WebKit &  6.9 & 1.1 & 2.21\\
Chromium & 6.6 & 1.1 & 2.06\\ 
64-bit & 33 & 0.70 & 6.62\\
\bottomrule\restartrowcolors{}
\end{tabular}
}
\end{table}






\section{Conclusion}

Our experiments suggest that both the WebKit and Chromium
approaches are several times faster than a naive approach (\texttt{string::find\_first\_of}) for the HTML scanning problem. 
We also find that an approach that indexes 64~bytes can be several times faster once again. However, our approach cannot serve as a simple drop-in replacement: it would require some reengineering to keep track of the indexes. Future work should investigate its application and document its  potential benefits.

The accelerated HTML parsing function is not the sole application of SIMD parsing in major Web engines. The WebKit engine has accelerated some functions having to do with JSON parsing, while both the Chromium and the WebKit engines are considering faster routines for CSS parsing. Future work should examine these algorithms. 

The HTML scanning problem is a specific task but it is generalizable. For example, the programmer might provide an arbitrary set of characters, and ask for a function to find all occurrences. Or there might be several classes of characters and we may want to have the ability to scan the text quickly while receiving information about the character types. In the HTML scanning scenario, the characters are part of the ASCII character set. The more general problem where the characters are arbitrary Unicode characters could prove more challenging. 

Future work should consider other processor architectures. Our approach based on blocks of 64~bytes should be even more favorable on processors with wider SIMD registers (e.g., AVX-512). Some instruction sets (RISC-V and SVE/SVE2) have registers of sizes that are potentially unknown at compile-time. We might need new algorithms for such instruction sets.






\bibliography{simdhtml}

\end{document}